\documentclass[12pt,a4paper]{article}
\usepackage{amsmath,amssymb,amsthm}
\usepackage{graphicx}
\usepackage{cite}

\renewcommand{\baselinestretch}{1.5}

\begin{document}

\title{{\Large \textbf{Low-frequency Magneto-optical Spectra of Bilayer
Bernal Graphene}}\\
}
\author{ Y. H. Ho$^{1,2}$, {Y. H. Chiu$^{1,\ast }$}~, D. H. Lin$^{2}$, C. P.
Chang$^{3}$, and M. F. Lin$^{1,\dagger }$ \\
{\footnotesize $^{1}$Department of Physics, National Cheng Kung University,
Tainan, Taiwan 701}\\
{\footnotesize $^{2}$Department of Physics, National Sun Yat-Sen University,
Kaohsiung, Taiwan 804}\\
{\footnotesize $^{3}$Center for General Education, Tainan University of
Technology, 710 Tainan, Taiwan}\\
}
\date{\today}
\maketitle

\begin{abstract}
The low-frequency magneto-optical absorption spectra of bilayer Bernal
graphene are studied within the tight-binding model and gradient
approximation. The interlayer interactions strongly affect the electronic
properties of the Landau levels (LL's), and thus enrich the optical
absorption spectra. According to the characteristics of the wave functions,
the low-energy LL's can be divided into two groups. This division results in
four kinds of optical absorption peaks with complex optical selection rules.
Observing the experimental convergent absorption frequencies close to zero
magnetic field might be useful and reliable in determining the values of
several hopping integrals. The dependence of the optical absorption spectra
on the field strength is investigated in detail, and the results differ
considerably from those of monolayer graphene. \vskip0.5 truecm


\noindent \textbf{\textrm{PACS numbers:}} {78.20.Ls, 78.67.Pt, 73.20.At,
71.15.Dx} \pagebreak
\end{abstract}


\renewcommand{\baselinestretch}{1.4}

\newpage \renewcommand{\baselinestretch}{1.5}

Since few-layer graphenes were produced by the mechanical friction$^{1,2}$
and thermal decomposition,$^{3,4}$ they have attracted considerable
experimental and theoretical studies. Monolayer graphene (MG) and bilayer
Bernal graphene (BBG; AB-stacked bilayer graphene), two of the simplest
few-layer graphenes, display very different electronic features. MG exhibits
linear energy dispersions near the Fermi level $E_{F}=0$ owing to the
hexagonal structure. However, BBG possesses two pairs of parabolic bands$%
^{5,6}$ owing to the interlayer interactions and stacking configurations.$%
^{7-9}$ These unusual electronic structures of few-layer graphenes have been
experimentally verified through angular-resolved photoemission spectroscopy.$%
^{10,11}$

In the presence of a uniform perpendicular magnetic field $\mathbf{B}=B_{0}%
\hat{z}$, the magnetoelectronic properties of few-layer graphenes have been
obtained by measurements of the quantum Hall conductivity,$^{2,12,13}$
scanning tunneling spectroscopy,$^{14,15}$ and cyclotron resonance.$^{16,17}$
The theoretical results show that the linear bands of MG\ become the Landau
levels (LL's) characterized by $E_{n}^{c,v}\propto \sqrt{n^{c,v}B_{0}}$,$%
^{18,19}$ where $c$ ($v$) is the index of the conduction (valence) LL's. $%
E_{n}^{c.v}$ and $n^{c,v}$ are the energy and the integer quantum number of
the $n$th energy band, respectively. The optical absorption peaks possess
the selection rule $\left\vert \triangle n\right\vert =\left\vert
n^{c}-n^{v}\right\vert =1$ and their frequency is proportional to $\sqrt{%
B_{0}}$,$^{20}$ which has been confirmed by magnetotransmission measurements.%
$^{16,21,22}$ For BBG, the effective-mass approximation, which neglects some
interlayer interactions, predicts that the magnetic field changes the
parabolic bands into LL's following $E_{n}^{c,v}\propto \sqrt{%
n^{c,v}(n^{c,v}-1)}B_{0}$.$^{20,23,24}$ The optical absorption frequencies
obey a linear function of $B_{0}$ in weak fields, and depend linearly on $%
\sqrt{B_{0}}$ as the corresponding energy leaves the parabolic band region.$%
^{20}$ The optical selection rule related to low-lying LL's is $\left\vert
\triangle n\right\vert =1$,$^{23}$ the same rule that is applied to MG.
However, the calculations of Peierls tight-binding model including more
important interlayer interactions indicate that the low-energy LL's should
exhibit more complex features,$^{24}$ i.e., there exists an energy gap,
asymmetry of LL's, two groups of LL's ($^{1st}$LL's and $^{2nd}$LL's), and
composite relations between $E_{n}^{c,v}$ and $B_{0}$. The varied electronic
features are reflected in the magneto-optical properties.

In this work, the influence of the important interlayer interactions on the
magneto-optical absorption spectra of BBG is investigated in detail by
gradient approximation$^{25,26}$ and Peierls tight-binding model (see the
geometry and detailed electronic properties in Ref. 24). The band-like matrix%
$^{24,27-31}$ is designed to solve a very huge Hamiltonian matrix. In this
way, the wave functions$^{24}$ are obtained and can be used to divide the
LL's of BBG into two groups $^{1st}$LL's and $^{2nd}$LL's. For the
magneto-optical excitations at a small magnetic field, the matrix will be
larger and more complicated in the calculations, which makes the computation
time very long. By utilizing the band-like matrix and obtaining the
localized feature of wave functions, we can efficiently reduce the
calculation time and thus evaluate the magneto-optical absorption spectra.
The absorption peaks exhibit complex selection rules and composite
field-dependent absorption frequencies. What deserves to be mentioned is
that our study is more comprehensive than other studies in optical
properties. The previous experimental and theoretical works mainly focused
on the optical properties owing to $^{1st}$LL's. However, in our work, all
of the possible optical absorption peaks resulting from the two groups are
discussed in detail, and could be clearly distinguished from each other.
Furthermore, a comparison of magneto-optical absorption spectra of MG and
BBG is also made.

The dispersionless LL's of BBG at $B_{0}=40$ T are shown in Fig. 1(a). The
interlayer interactions strongly influence the main features of LL's. The
conduction and valence bands are asymmetric. There is an energy gap between
the lowest unoccupied LL and the highest occupied LL. Based on the
characteristics of wave functions, LL's can be classified into two groups, $%
^{1st}$LL's (black curves) and $^{2nd}$LL's (red curves). $^{1st}$LL's
appears at $\left\vert E^{c,v}\right\vert >0$, and $^{2nd}$LL's begins at $%
E^{c}\gtrsim \gamma _{1}+\gamma _{6}$ (0.34 eV) and $E^{v}\lesssim -\gamma
_{1}+\gamma _{6}$ (-0.39 eV). $\gamma _{i}$'s$^{6,24,32}$ are the most
important hopping integrals. The energies of these LL's exhibit the
composite relationship of $E_{n}^{c,v}\propto B_{0}$ and $E_{n}^{c,v}\propto
\sqrt{B_{0}}$.$^{24}$

The characteristics of wave functions deserve a closer investigation because
they strongly dominate the optical properties. Each LL is composed of
fourfold degenerate states with similar characteristics. At $k_{x}=k_{y}=0$,
the wave functions mainly consist of four localized subenvelope functions
centered at positions $x_{1}=1/6$, $x_{2}=1/3$, $x_{3}=2/3$, and $x_{4}=5/6$%
. $x_{j=1-4}=A_{i,N}/2R_{B}$ or $=B_{i,N}/2R_{B}$, where $i$ is the index of
the $i$th layer and $N$ represents the $N$th $A_{i}$ or $B_{i}$ atoms. $%
R_{B}^{24,27}$ defines the dimensionality of the Hamiltonian matrix. The
feature of wave functions resulting from the atoms with an even number ($%
A_{i,e}^{c,v}$ and $B_{i,e}^{c,v}$) is similar to that with an odd ($%
A_{i,o}^{c,v}$ and $B_{i,o}^{c,v}$) number. Because of this, only odd atoms
are discussed in following calculations, where $e$ ($o$) is an even (odd)
integer. For convenience, one of the fourfold degenerate states at $%
x_{1}=1/6 $ is selected to be examined, as shown in Figs. 1(b)-1(e). Through
appropriate fitting, the wave functions of the $n_{1}$th $^{1st}$LL (black
curves) and $n_{2}$th $^{2nd}$LL (red curves) can be expressed as
\begin{subequations}
\begin{align}
A_{1,o}^{c}\text{, }A_{2,o}^{c}& \propto \varphi _{2}\left( x\right) \text{,
}B_{1,o}^{c}\propto \varphi _{1}\left( x\right) \text{, }B_{2,o}^{c}\propto
\varphi _{0}\left( x\right) \text{, }n_{1}^{c,eff}=0\text{ for }n_{1}=1\text{%
;} \\
A_{1,o}^{v}\text{, }A_{2,o}^{v}& \propto \varphi _{0}\left( x\right) \text{,
}B_{1,o}^{v}\propto \varphi _{2}\left( x\right) \text{, }B_{2,o}^{v}\propto
\varphi _{1}\left( x\right) \text{, }n_{1}^{v,eff}=1\text{ for }n_{1}=1\text{%
;} \\
A_{1,o}^{c,v}\text{, }A_{2,o}^{c,v}& \propto \varphi _{n_{1}-1}\left(
x\right) \text{, }B_{1,o}^{c,v}\propto \varphi _{n_{1}-2}\left( x\right)
\text{, }B_{2,o}^{c,v}\propto \varphi _{n_{1}}\left( x\right) \text{ for }%
n_{1}\text{ (}=n_{1}^{c,v,eff}\text{)}\geq 2\text{.}
\end{align}%
\end{subequations}
\begin{subequations}
\begin{align}
A_{1,o}^{c,v}\text{, }A_{2,o}^{c,v}& \propto \varphi _{0}\left( x\right)
\text{, }B_{1,o}^{c,v}\propto \varphi _{2}\left( x\right) \text{, }%
B_{2,o}^{c,v}\propto \varphi _{1}\left( x\right) \text{, }n_{2}^{c,v,eff}=0%
\text{ for }n_{2}=1\text{;} \\
A_{1,o}^{c,v}\text{, }A_{2,o}^{c,v}& \propto \varphi _{n_{2}-1}\left(
x\right) \text{, }B_{1,o}^{c,v}\propto \varphi _{n_{2}-2}\left( x\right)
\text{, }B_{2,o}^{c,v}\propto \varphi _{n_{2}}\left( x\right) \text{ for }%
n_{2}\text{ (}=n_{2}^{c,v,eff}+1\text{)}\geq 2\text{.}
\end{align}%
$\varphi _{n}\left( x\right) $ is the product of the $n$th-order Hermite
polynomial and Gaussian function,$^{20}$ where $n$ is the number of zero
points in $\varphi _{n}\left( x\right) $ and chosen to define the quantum
number of wave functions.$^{24}$ $n_{1}^{c,v,eff}$ ($n_{2}^{c,v,eff}$) is
the effective quantum number of the $n_{1}$th $^{1st}$LL's ($n_{2}$th $%
^{2nd} $LL's) defined by the quantum number of $B_{2,o}^{c,v}$ ( $%
A_{1,o}^{c,v}$). It should be noted that the $^{1st}$LL's and $^{2nd}$LL's
could be distinguished by the different features of the wave functions in
the mixed region ($E^{c}\gtrsim 0.34$ eV and $E^{v}\lesssim -0.39$ eV),
which is useful in analyzing the optical excitation channels.

When BBG is subjected to an electromagnetic field at zero temperature, only
excitations from the valence to the conduction bands occur. The optical
selection rule $\Delta \mathbf{k=0}$ due to the vertical transitions is
mainly determined by the zero momentum of the photon. Based on the Fermi's
golden rule, the optical absorption function is given by
\end{subequations}
\begin{eqnarray}
A(\omega ) &\propto &\sum_{n_{1,2}^{c,eff},n_{1,2}^{v,eff}}\int_{1stBZ}{%
\frac{d\mathbf{k}}{(2\pi )^{2}}}|\langle \psi ^{c}(n_{1,2}^{c,eff},\mathbf{k}%
)|{\frac{\widehat{\mathbf{E}}\cdot \mathbf{P}}{m_{e}}}|\psi
^{v}(n_{1,2}^{v,eff},\mathbf{k})\rangle |^{2}  \notag \\
&&\times Im\{\frac{f\left[ E^{c}\left( \mathbf{k},n_{1,2}^{c,eff}\right) %
\right] -f\left[ E^{v}\left( \mathbf{k},n_{1,2}^{v,eff}\right) \right] }{%
E^{c}\left( \mathbf{k},n_{1,2}^{c,eff}\right) -E^{v}\left( \mathbf{k}%
,n_{1,2}^{v,eff}\right) -\omega -i\Gamma }\}\text{,}
\end{eqnarray}%
where $\widehat{\mathbf{E}}$ is the unit vector of an electric polarization.
The velocity matrix element $M^{cv}=\langle \psi ^{c}(n_{1,2}^{c},\mathbf{k}%
)|\widehat{\mathbf{E}}\cdot \mathbf{P/}m_{e}|\psi ^{v}(n_{1,2}^{v},\mathbf{k}%
)\rangle $ is calculated within the gradient approximation. $M^{cv}$\
strongly depends on the characteristics of wave functions and would dominate
the peak intensity. Due to the orthogonality of $\varphi _{n}(x)$, $M^{cv}$
has nonzero values except one condition when $A_{i,o}^{c,v}$ ($A_{i,e}^{c,v}$%
) and $B_{i,o}^{c,v}$ ($B_{i,e}^{c,v}$) own the same $\varphi _{n}(x)$.
Accordingly, the selection rules of optical absorption peaks can easily be
obtained by the definition of wave functions in Eqs. (1) and (2).

The low-frequency optical absorption spectrum of BBG at $B_{0}=40$ T
presents many prominent and some inconspicuous (indicated by arrows) peaks,
as shown in Fig. 2(a). The prominent peaks with definite optical selection
rules are discussed in the following paragraphs. The peaks can mainly be
classified into four kinds of peaks, $\omega _{11}$'s (black dots), $\omega
_{22}$'s (red dots), $\omega _{12}$'s (green dots), and $\omega _{21}$'s
(blue dots), which originate in the excitations of $^{1st}$LL$^{v}$s$%
\rightarrow ^{1st}$LL$^{c}$s, $^{2nd}$LL$^{v}$s$\rightarrow ^{2nd}$LL$^{c}$%
s, $^{1st}$LL$^{v}$s$\rightarrow ^{2nd}$LL$^{c}$s, and $^{2nd}$LL$^{v}$s$%
\rightarrow ^{1st}$LL$^{c}$s, respectively. The former (latter) two kinds of
peaks result from the transitions between two LL's in the same (different)
groups, as illustrated by the inset of Fig. 2(d). The four kinds of
absorption peaks might have different optical selection rules because of the
characteristics of wave functions in the two groups of LL's.

The optical excitations of each prominent peak can clearly be identified, as
shown in Fig. 2(a). For convenience, the excitations of $\omega _{11}$'s, $%
\omega _{22}$'s, $\omega _{12}$'s, and $\omega _{21}$'s are, respectively,
represented as $n_{1}^{v,eff}\rightarrow n_{1}^{c,eff}$, $%
n_{2}^{v,eff}\rightarrow n_{2}^{c,eff}$, $n_{1}^{v,eff}\rightarrow
n_{2}^{c,eff}$, and $n_{2}^{v,eff}\rightarrow n_{1}^{c,eff}$ in the
following. The selection rule of $\omega _{ij}$'s is denoted by $\triangle
n_{ii^{\prime }}$ ($=n_{i^{\prime }}^{c,eff}-n_{i}^{v,eff}$). As to $\omega
_{11}$'s and $\omega _{22}$'s, only the first peak ($\omega _{11}^{1}$) of $%
\omega _{11}$'s originates in single channel, $1\rightarrow 2$. The other $m$%
th peak $\omega _{11}^{m}$ ($\omega _{22}^{m}$) consists of the pair $\omega
_{11}^{m,L}$ and $\omega _{11}^{m,H}$ ($\omega _{22}^{m,L}$ and $\omega
_{22}^{m,H}$) corresponding to $m\rightarrow m+1$ ($m-1\rightarrow m$) and $%
m+1\rightarrow m$ ($m\rightarrow m-1$), respectively. The superscript $L$ ($%
H $) indicates the lower (higher) energy of the pair. For example, $\omega
_{11}^{2}$ ($\omega _{22}^{1}$) is composed of $\omega _{11}^{2,L}$ and $%
\omega _{11}^{2,R}$ ($\omega _{22}^{1,L}$ and $\omega _{22}^{1,R}$) owing to
$2\rightarrow 3$ and $3\rightarrow 2$ ($0\rightarrow 1$ and $1\rightarrow 0$%
), respectively. The reason for this is that the energies of $m\rightarrow
m+1$ ($m-1\rightarrow m$) and $m+1\rightarrow m$ ($m\rightarrow m-1$) in $%
\omega _{11}$'s ($\omega _{22}$'s) are slightly different owing to the
asymmetry of the conduction and valence LL's. In short, the optical
selection rules of $\omega _{11}$'s and $\omega _{22}$'s can be represented
by $\triangle n_{11}=\triangle n_{22}=\pm 1$, which is the same for the LL's
in MG.

As for the transitions between two different groups, the $m$th peak of $%
\omega _{12}$'s is formed with the pair, $\omega _{12}^{m,L}$ originating in
$m\rightarrow m$ and $\omega _{12}^{m,H}$ resulting from $m+1\rightarrow m-1$%
. For instance, the first pair is indicated by $\omega _{12}^{1,L}$ and $%
\omega _{12}^{1,H}$ in Fig. 2(a), which correspond to $1\rightarrow 1$ and $%
2\rightarrow 0$, respectively. That is to say, $\omega _{12}^{m,L}$ owns the
optical selection rule $\triangle n_{12}=0$ and $\omega _{12}^{m,H}$
possesses $\triangle n_{12}=-2$. For $\omega _{21}$'s, the excitations $%
0\rightarrow 0$ and $0\rightarrow 2$ lead to the peaks $\omega _{21}^{1,L}$
and $\omega _{21}^{1,H}$ of the first pair in Fig. 2(a), respectively. The
other channels $m-1\rightarrow m+1$ and $m\rightarrow m$ result in the pair $%
\omega _{21}^{m,L}$ and $\omega _{21}^{m,H}$, respectively. In other words,
the selection rules of $\omega _{21}$'s are $\triangle n_{21}=0$ and $%
\triangle n_{21}=2$. Obviously, the optical selection rules of $\omega _{12}$%
's and $\omega _{21}$'s are different from those of $\omega _{11}$'s and $%
\omega _{22}$'s, which is not clearly discussed in the previous theoretical
works.

The absorption rate significantly relies on the field strength and the
excitation energy. In Figs. 2(a)-2(d), the peak intensity augments with
increasing field strength, while the opposite is true for the peak number.
Furthermore, some peaks with commensurate energies are overlapping because
the peak spacing declines as the frequency increases, which varies the peak
intensity. In contrast to BBG, MG exhibits absorption peaks with uniform
intensity (Fig. 2(e)). The interlayer interactions are the main reasons for
the difference between BBG and MG. The field-dependent absorption
frequencies ($\omega _{a}$'s) of $\omega _{11}$'s, $\omega _{22}$'s, $\omega
_{12}$'s and $\omega _{21}$'s are shown in Fig. 3(a) by the black, red,
green, and blue dots, respectively. $\omega _{a}$'s rise with increasing $%
B_{0}$ and might follow the composite relationship of $\omega _{a}\propto
B_{0}$ and $\omega _{a}\propto \sqrt{B_{0}}$. Apparently, $\omega
_{a}\propto \sqrt{B_{0}}$ in MG (Fig. 3(b)) is very different from that in
BBG. In addition, the convergent frequencies of $\omega _{11}$'s, $\omega
_{22}$'s, $\omega _{12}$'s and $\omega _{21}$'s at $B_{0}\rightarrow 0$ are
approximately 0, 0.73 eV (2$\gamma _{1}$), 0.34 eV ($\gamma _{1}+\gamma _{6}$%
), and 0.39 eV ($\gamma _{1}-\gamma _{6}$), respectively. This implies that
the optical measurements can reasonably determine the values of $\gamma _{1}$
and $\gamma _{6}$ through observing the convergent frequencies of absorption
peaks. The predicted result is very useful and reliable for experimental
studies.

The tight-binding model is widely applied to tackle the physical problems of
semiconductors and carbon-related systems. Through the band-like matrix and
the localized feature of wave functions, the huge and complicated
Hamiltonian matrix can be further simplified in the calculations of the
magneto-optical excitations for small field cases. Thus the computation time
is substantially reduced. Moreover, the numerical strategy can be utilized
to study not only the electronic and optical properties, but also the
Coulomb excitations and transport properties in other crystal materials.

In summary, it can be said that the interlayer interactions strongly affect
the magneto-optical properties of BBG in the presence of uniform
perpendicular magnetic fields. Based on the characteristics of wave
functions, the conduction and valence LL's are asymmetric and can be divided
into $^{1st}$LL's and $^{2nd}$LL's leading to four kinds of optical
absorption peaks $\omega _{11}$'s, $\omega _{22}$'s, $\omega _{12}$'s, and $%
\omega _{22}$'s. Their optical selection rules are $\triangle n_{11}=\pm 1$,
$\triangle n_{22}=\pm 1$, $\triangle n_{12}=0,-2$, and $\triangle n_{21}=0,2$%
, respectively, which can again be explained by the characteristics of wave
functions. The selection rules and field-dependent absorptions of BBG are
more complicated than those of MG. The above-mentioned influences of
interlayer interactions on the magneto-optical properties could be confirmed
by magnetoabsorption spectroscopy measurements.$^{17,21,22}$ Furthermore,
the convergent absorption frequencies at nearly zero field might be helpful
and reliable in determining the interlayer atomic interactions $\gamma _{1}$
and $\gamma _{6}$ through the magneto-optical measurements.

\vskip 0.6 truecm

\noindent \textbf{Acknowledgments} \vskip 0.3 truecm

This work was supported by the NSC and NCTS of Taiwan, under the grant No.
NSC 95-2112-M-006-028-MY3 and NSC 97-2112-M-110-001-MY2.

\newpage $^{\ast }$Electronic address: airegg.py90g@nctu.edu.tw

$^{\dagger }$Electronic address: mflin@mail.ncku.edu.tw (corresponding
author to whom should be addressed)

\begin{itemize}
\item[$^{1}$] K. S. Novoselov \textit{et al.}, Science \textbf{306}, 666
(2004).

\item[$^{2}$] K. S. Novoselov \textit{et al.}, Nature \textbf{438}, 197
(2005).

\item[$^{3}$] C. Berger \textit{et al.}, Science \textbf{312}, 1191 (2006).

\item[$^{4}$] E. Stolyarova \textit{et al.}, PNAS \textbf{104}, 9209 (2007).

\item[$^{5}$] J. C. Slonczewski and P. R. Weiss, Phys. Rev\textit{.} \textbf{%
109}, 272 (1958).

\item[$^{6}$] J.-C. Charlier and J.-P. Michenaud, Phys. Rev. B \textbf{46},
4531 (1992).

\item[$^{7}$] S. Latil and L. Henrard, Phys. Rev. Lett. \textbf{97}, 036803
(2006).

\item[$^{8}$] B. Partoens and F. M. Peeters, Phys. Rev. B \textbf{74},
075404 (2006).

\item[$^{9}$] C. L. Lu \textit{et al.}, Phys. Rev. B \textbf{73}, 144427
(2006).

\item[$^{10}$] T. Ohta \textit{et al.}, Science \textbf{313}, 951 (2006).

\item[$^{11}$] T. Ohta \textit{et al.}, Phys. Rev. Lett. \textbf{98}, 206802
(2007).

\item[$^{12}$] K. S. Novoselov \textit{et al.}, Nat. Phys\textit{.} \textbf{2%
}, 177 (2006).

\item[$^{13}$] Y. Zhang \textit{et al.}, Nature \textbf{438}, 201 (2005).

\item[$^{14}$] G. Li and E. Y. Andrei, Nat. Phys. \textbf{3}, 623 (2007).

\item[$^{15}$] T. Matsui \textit{et al.}, Phys. Rev. Lett. \textbf{94},
226403 (2005).

\item[$^{16}$] R. S. Deacon \textit{et al.}, Phys. Rev. B \textbf{76},
081406 (2007).

\item[$^{17}$] E. A. Henriksen \textit{et al.}, Phys. Rev. Lett. \textbf{100}%
, 087403 (2008).

\item[$^{18}$] J. W. McClure, Phys. Rev. \textbf{104}, 666 (1956).

\item[$^{19}$] Y. Zheng and T. Ando, Phys. Rev. B \textbf{65}, 245420 (2002).

\item[$^{20}$] M. Koshino and T. Ando, Phys. Rev. B \textbf{77}, 115313
(2008).

\item[$^{21}$] M. L. Sadowski \textit{et al.}, Phys. Rev. Lett. \textbf{97},
266405 (2006).

\item[$^{22}$] P. Plochocka \textit{et al.}, Phys. Rev. Lett. \textbf{100},
087401 (2008).

\item[$^{23}$] D. S. L. Abergel and V. I. Fal'ko, Phys. Rev. B \textbf{75},
155430 (2007).

\item[$^{24}$] Y. H. Lai \textit{et al.}, Phys. Rev. B \textbf{77}, 085426
(2008).

\item[$^{25}$] M. F. Lin and Kenneth W.-K. Shung, Phys. Rev. B \textbf{50},
17744 (1994).

\item[$^{26}$] C. P. Chang \textit{et al.}, Carbon \textbf{42}, 2975 (2004).

\item[$^{27}$] Y. H. Chiu \textit{et al.}, Phys. Rev. B \textbf{77}, 045407
(2008).

\item[$^{28}$] J. H. Ho \textit{et al.}, Nanotechnology \textbf{19}, 035712
(2008).

\item[$^{29}$] Y. C. Huang \textit{et al.}, Nanotechnology \textbf{18},
495401 (2007).

\item[$^{30}$] Y. H. Chiu \textit{et al.}, Phys. Rev. B \textbf{78}, 245411
(2008).

\item[$^{31}$] Y. C. Huang \textit{et al.}, J. Appl. Phys. \textbf{103},
073709 (2008).

\item[$^{32}$] J.-C. Charlier \textit{et al.}, Phys. Rev. B \textbf{43},
4579 (1993).
\end{itemize}

\newpage \centerline {\Large \textbf {Figure Captions}}

\vskip0.3 truecm

FIG. 1. (a) Landau levels of bilayer Bernal graphene at $B_{0}$=40 T. The
wave functions of (b) $A_{1,o}$, (c) $B_{1,o}$, (d) $A_{2,o}$, (e) $B_{2,o}$
atoms with odd integer indices are shown.

\vskip0.5 truecm

FIG. 2. The optical absorption spectra of bilayer Bernal graphene at (a) 40
T, (b) 30 T, (c) 20 T, and (d) 10 T. The spectrum of monolayer graphene at
40 T is plotted in (e).

\vskip0.5 truecm

FIG. 3. The field-dependent optical absorption frequencies of (a) bilayer
Bernal graphene and (b) of monolayer graphene.

\end{document}